\documentclass[aps,prb,twocolumn,groupedaddress,showpacs,intlimits,amsmath,amssymb,floatfix,citeautoscript,footinbib]{revtex4-1}

\usepackage[final]{graphicx}
\usepackage[hang,center]{subfigure}
\usepackage{amssymb}
\usepackage{amsmath}

\emergencystretch=\maxdimen
\newcommand{\cto}{Ca$_3$Ti$_2$O$_7$}

\newcommand{\atwo}{$A2_1am$}
\newcommand{\ifour}{$I4/mmm$}
\newcommand{\xthree}{$X_3^-$}
\newcommand{\xtwo}{$X_2^+$}
\newcommand{\gfive}{$\Gamma_5^-$}

\newcommand{\ntho}{$\eta_1^{X_3^-}$}
\newcommand{\ntht}{$\eta_2^{X_3^-}$}
\newcommand{\ntwo}{$\eta_1^{X_2^+}$}
\newcommand{\ntwt}{$\eta_2^{X_2^+}$}
\newcommand{\npo}{$\eta_1^{P}$}
\newcommand{\npt}{$\eta_2^{P}$}
\newcommand{\napo}{$\eta_1^{AP}$}
\newcommand{\napt}{$\eta_2^{AP}$}
\newcommand{\yo}{$Z_1$}
\newcommand{\yt}{$Z_2$}
\newcommand{\zo}{$Z_3$}
\newcommand{\zt}{$Z_4$}

\begin{document}
\title{Domains and ferroelectric switching pathways in Ca$_3$Ti$_2$O$_7$ from first principles}
\author{Elizabeth A. Nowadnick}
\email{E-mail: nowadnick@cornell.edu}
\author{Craig J. Fennie}
\affiliation{School of Applied and Engineering Physics, Cornell University, Ithaca, NY 14853, USA}

\date{\today}


\begin{abstract}
Hybrid improper ferroelectricity, where an electrical polarization can be induced via a trilinear coupling to two non-polar structural distortions of different symmetry, has recently been experimentally demonstrated for the first time in the $n$=2 Ruddlesden-Popper compound \cto.  In this paper we use  group theoretic methods and first-principles calculations to identify possible ferroelectric switching pathways in \cto. We identify low-energy paths that reverse the polarization direction  by switching via an orthorhombic twin domain, or via an antipolar structure. We also introduce a chemically intuitive set of local order parameters to give insight into how  these paths are relevant to switching nucleated at domain walls. 
Our findings suggest that switching may proceed via more than one mechanism in this material.     
\end{abstract}

\maketitle

\section{Introduction}
Ferroelectric materials display a broad range of fascinating properties, from their phase transition and  switching behaviors,~\cite{rabe-book} to various photoinduced effects~\cite{kreisel2012} and intriguing topological defects in their domain structures.~\cite{choi10,yadav2016}
They also have  utility in applications such as low-power electronics.~\cite{scott07}
In particular,  there is a great deal of interest in ferroelectrics that allow a coupling between the polarization and another order parameter (OP), making  electric field control of non-polar OPs possible. 
 For example,  the search for strategies to directly couple  magnetization and polarization provided an impetus to understand new mechanisms for ferroelectricity,~\cite{birol2012} such as spin-induced ferroelectricity,~\cite{kimura03, tokura14} and more recently, octahedral rotation-induced ferroelectricity.~\cite{bousquet08,benedek11,rondinelli12a, benedek15}
This latter type of ferroelectricity, termed ``hybrid improper ferroelectricity,"  is a mechanism where a polarization can be induced via a trilinear coupling to two  octahedral rotations (or other structural distortions~\cite{varignon15}) of different symmetry. Hybrid improper ferroelectricity was predicted theoretically in ABO$_3$/A$^\prime$BO$_3$  superlattices~\cite{bousquet08, rondinelli12a,zhao14,varignon15,xu2015} and in $n$=2 Ruddlesden-Popper (RP) materials.~\cite{benedek11} Recently, it was experimentally demonstrated in the $n$=2 RP compound \cto,~\cite{oh15} and a complex domain structure was observed.~\cite{huang2016,huang2016-2}  The observation of an unexpectedly low switching barrier and abundant structural domains suggests that these domains may be critical to ferroelectric switching, but the precise pathway by which the polarization reverses remains an open question.

This paper aims to take the first steps to address this question. While a rigorous theoretical description of the full dynamic ferroelectric switching  process is challenging, simpler approaches can provide valuable insight, as has been demonstrated by previous work on BiFeO$_3$~\cite{heron14}. In this work, we first  use a combination of group theory and first-principles calculations to survey the energetics of Ca$_3$Ti$_2$O$_7$ in a space of low-energy metastable structures, and enumerate the possible ferroelectric switching pathways within this space. Then, we introduce a set of OPs  that reflect the structural chemistry of  Ca$_3$Ti$_2$O$_7$ to aid in understanding how these pathways are relevant for switching nucleated at domain walls. 
While this paper focuses on Ca$_3$Ti$_2$O$_7$, the approach we introduce  is generic to all A$_3$B$_2$O$_7$ RP materials.
 
\begin{figure*}
\begin{center}
 \includegraphics[width=0.9\textwidth]{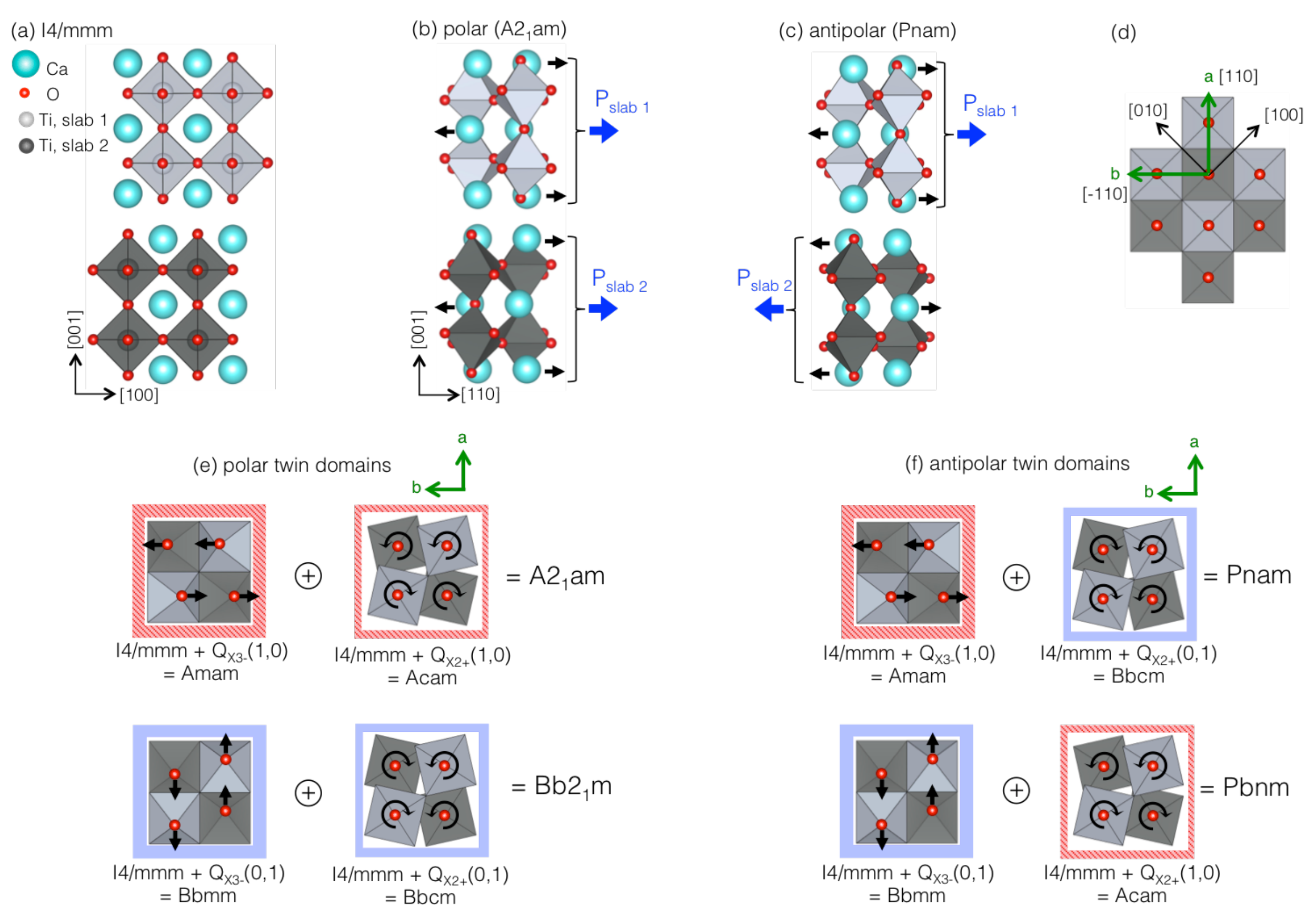}
 \caption{\label{struc} Ca$_3$Ti$_2$O$_7$ structure and orthorhombic twin domains: (a) high-symmetry structure $I4/mmm$, (b)  polar structure  $A2_1am$ and (c) antipolar structure  $Pnam$. The upper (lower) $n$=2 CaTiO$_3$  slab is indicated by light (dark) grey; in (b, c) the blue arrows give the  polarization direction in each perovskite slab. (d) The orientation of the orthorhombic axes ${\bf a}$ and ${\bf b}$ relative to the tetragonal axes, which determines the setting of the  orthorhombic space group symbols.  
 (e) The octahedral rotation distortions that transform like the $(a,0)$ (top, pink hatched) and $(0,a)$ (bottom, light blue) directions of irreps $X_3^-$ and $X_2^+$, respectively. Taking both irreps along $(a,0)$ leads to  polar $A2_1am$, while taking both along $(0,a)$ leads to polar orthorhombic twin $Bb2_1m$. (f) The combinations of $X_3^-$ and $X_2^+$ irrep directions that lead to the two orthorhombic twins of the antipolar structure, $Pnam$ and $Pbnm$ (top and bottom).
Ca ions and planar oxygens are suppressed in (d-f) for clarity.}
 \end{center}
 \end{figure*}

Fig.~\ref{struc}(a) shows the high-symmetry parent  structure $I4/mmm$, which consists of CaO-terminated CaTiO$_3$  slabs of thickness $n$=2, separated by a rocksalt layer. Adjacent perovskite slabs are offset from each other by $a_0$/2[110], where $a_0$ is the lattice constant.  At  room temperature \cto~crystallizes in the orthorhombic polar space group \atwo.~\cite{elcombe91,senn15} 
This distorted structure can be decomposed into three distinct structural distortions that transform like irreducible representations (irreps) of $I4/mmm$:~\cite{benedek11} two octahedral rotation-like distortions  that transform like $X_3^-$ and $X_2^+$, respectively, and a polar distortion that transforms like $\Gamma_5^-$.   

The $A2_1am$ symmetry is established by $X_3^-\oplus X_2^+$, the polar distortion is not required. 
The polar distortion is induced because its OP $Q_P$ and the  `hybrid' OP $Q_{X_3^-} Q_{X_2^+}$ transform in the same way under the symmetry operations of $I4/mmm$, so a trilinear coupling between these OPs is allowed 
 in the Landau expansion of the energy:~\cite{benedek11}
\begin{equation}
\mathcal{F}_\mathrm{tri} =\alpha Q_{X_3^-} Q_{X_2^+} Q_P
\label{eq:tri-1d}
\end{equation}
(here the OP $Q$ is the distortion amplitude). It is clear from Eq.~\ref{eq:tri-1d} that reversing the polarization direction ($Q_P \rightarrow -Q_P$) requires reversing  one but not both of the octahedral rotation OPs, thus, Ref.~\onlinecite{benedek11} identified two possible switching pathways. To reverse the polarization direction via either of these pathways, the amplitude of one of the octahedral rotations is brought to zero, and then the rotation is turned back on with the opposite sense.

What are other possible switching pathways? To make progress towards answering this question, in Fig.~\ref{struc}(b) we take a closer look at the polar distortion in $A2_1am$, which primarily consists of a 2-against-1 displacement of the Ca ions in each $n=2$ perovskite slab.~\cite{benedek12,mulder13} The atomic displacements in adjacent perovskite slabs are in the same direction, leading to a net polarization.  One may wonder, what would happen if each slab still has the same 2-against-1 displacement of Ca ions, but now in opposite directions, which would lead to an antipolar structure,  shown in Fig.~\ref{struc}(c)? This antipolar structure may be energetically close to the polar one, because the perovskite slabs are weakly connected across the rocksalt layer.
The same combination of irreps ($X_3^-\oplus X_2^+$) that induces the polar distortion in $A2_1am$ also can induce this antipolar distortion, when the irreps are taken along different directions in OP space. 
The key to describing these polar and antipolar structures on an equal footing is to use the full two-dimensional OPs~\cite{harris11,foot3} to describe the structural distortions in \cto. This means that both an amplitude $Q$ and a phase $\theta$ characterize each distortion. Note that Ref.~\onlinecite{benedek11} worked with OPs  restricted to one dimension ($Q$ only, neglected $\theta$); consideration of the full two-dimensional OPs leads to additional trilinear coupling terms beyond  Eq.~\ref{eq:tri-1d}, which we will discuss later.  
In this paper, we will use these two-dimensional OPs to first  describe the polar and antipolar structures, and their orthorhombic twin domains, and then systematically identify additional switching pathways beyond those mentioned above (for which the antipolar structure will be important). 

The plan for the rest of this paper is as follows. Sec.~\ref{methods} describes our methods. Sec.~\ref{sec-struc} introduces the necessary two-dimensional OP formalism, and uses these OPs to enumerate all domains of the polar and antipolar structures. In Sec.~\ref{section-switching}, we present the possible ferroelectric switching pathways for Ca$_3$Ti$_2$O$_7$. To understand how these pathways relate to the crystallographic structure and likely domain walls in the sample, we introduce local OPs in Sec.~\ref{section-local}, and discuss the implications of our results in Sec.~\ref{section-discussion}.
    
\begin{table*}[]
\begin{center}
\caption{\label{X3X2} Isotropy subgroups of \ifour~ generated by distinct directions of the irreps \xthree, \xtwo, and $X_3^- \oplus X_2^+$. Energies are given relative to that of the ground state \atwo. For isotropy subgroups $P2_1am$ and $P2_1nm$, where one of the OPs is along a general direction ($a,b$), the energy reported is that when $a=b$, obtained from a NEB calculation. }

\begin{tabular}{c | c | c |   c  | c |c | c | c | c|c|c|c}
\hline
\hline
irrep & ${\bf \eta}^{X_3^-}$ & ${\bf \eta}^{X_2^+}$ & space group & induced &${\bf P}$ & energy & $\eta^{X_3^-}$ & $\eta^{X_2^+}$ & space group & induced & ${\bf P}$ dir.\\
& & & & irrep & direction& [meV/Ti] & (twin) & (twin) & (twin) & irrep (twin) & (twin)\\
\hline

$X_3^-$& (a,0)  & -& $Amam$ & -&-& 56 & (0,a) & -& $Bbmm$  & -&-\\
& (a,a) & -& $P4_2/mnm$ & -&-& 45 & - & - & - & -&-\\
& (a,b) & -& $Pnnm$ &-& - & -&- & - & -& -&-\\
\hline
$X_2^+$& -& (a,0) & $Acam$ &-&-& 90 & -& (0,a) & $Bbcm$& -&-\\
& -& (a,a) & $P4/mbm$ &-&-& 172  & - & - & -& -&-\\
& -&  (a,b) & $Pbam$ &-& -&- & - & - & -& -&-\\
\hline

$X_3^-\oplus X_2^+$&(a,0) &(b,0) & $A2_1am$& $\eta^P_2$ &[110]& 0 &(0,a) &(0,b) & $Bb2_1m$ & $\eta^P_1$&[-110]\\
&(a,0) & (0,b) & $Pnam$ &$\eta^{AP}_2$&-& 7 &(0,a) & (b,0) & $Pbnm$ & $\eta^{AP}_1$&-\\
&(a,a) & (b,b) & $C2mm$ &$\eta^P_{1,2}$, $\eta^{AP}_{1,2}$ &[100]& 41 &(a,a) & (-b,b) & $Cm2m$ & $\eta^P_{1,2}$, $\eta^{AP}_{1,2}$&[010]\\
&(a,b) & (c,0) & $P2_1nm$ & $\eta^P_2$, $\eta^{AP}_1$ &[110]& 35&(a,b) & (0,c) & $Pn2_1m$ & $\eta^P_1$, $\eta^{AP}_2$&[-110]\\
&(a,0) & (b,c) & $P2_1am$ & $\eta^P_2$, $\eta^{AP}_2$ &[110]& 32&(0,a) & (b,c) & $Pb2_1m$  & $\eta^P_1$, $\eta^{AP}_1$&[-110]\\
&(a,b) & (c,d) & $Pm$ &$\eta^P_{1,2}$, $\eta^{AP}_{1,2}$& $a$[110]+& - & - & - & -& -\\
& &  &  && $b$[-110]&  &  &  & & \\
\hline

\end{tabular}
\end{center}
\end{table*}

\section{Methods \label{methods}}
We perform density functional theory calculations using  {\sc VASP}~\cite{kresse93, kresse99} with the PBEsol  functional~\cite{perdew08}, and use the nudged elastic band (NEB) method\cite{jonsson98} to compute switching paths. We use a $Z=4$ cell commensurate with both orthorhombic twins of Ca$_3$Ti$_2$O$_7$, a $6\times 6 \times 2$ $k$-point mesh, a 600 eV plane wave cutoff, and for structural relaxations a force convergence tolerance of 2 meV/\AA. 
We make use of {\sc ISOTROPY}~\cite{isotropy,campbell2006} to aid with the group theoretic analysis, and {\sc VESTA} to visualize crystal structures.~\cite{vesta}

\section{Crystallographic structure and domains \label{sec-struc}}

\begin{figure*}
\begin{center}
 \includegraphics[width=0.8\textwidth]{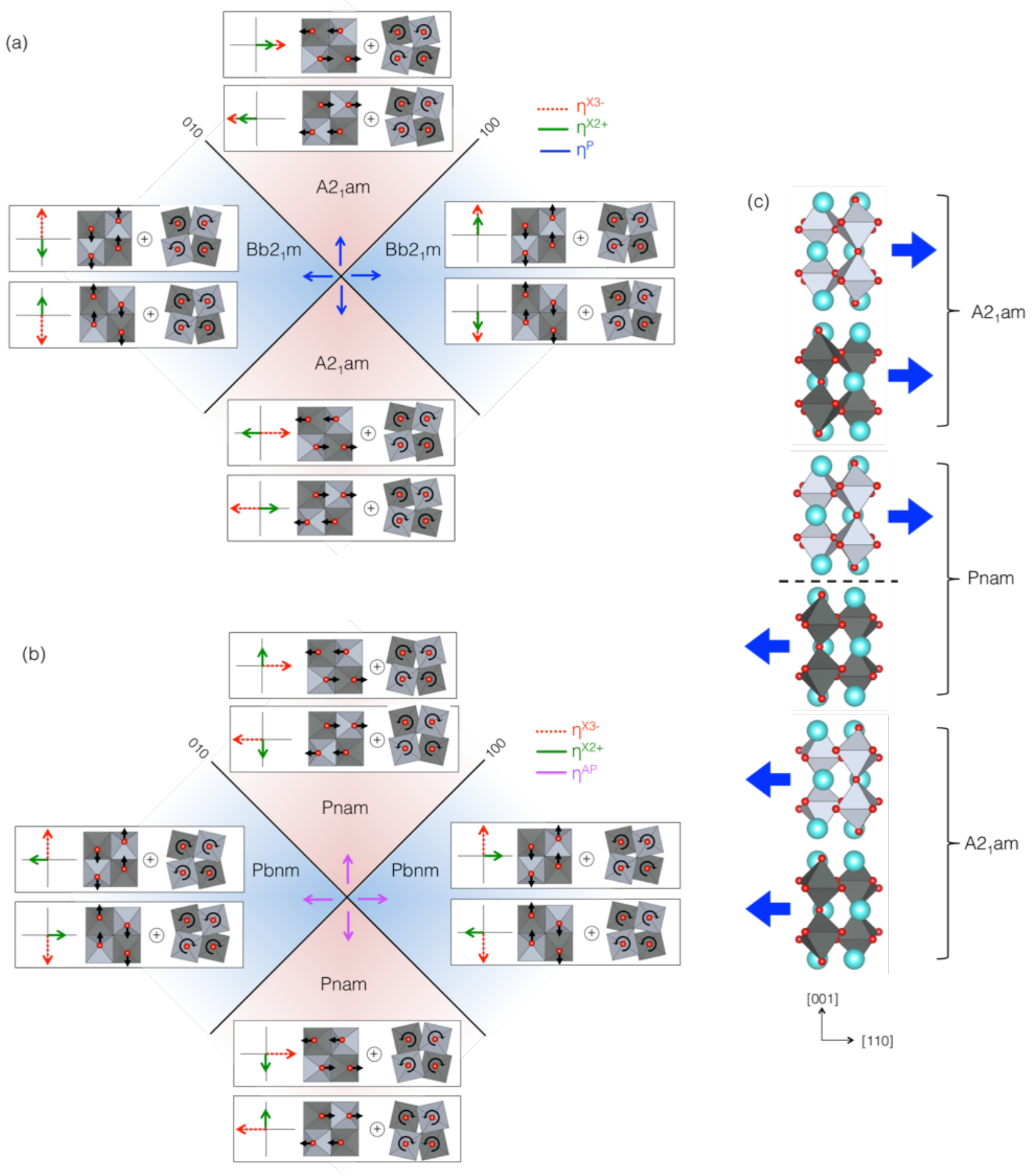}
 \caption{\label{domains} Eight structural domains of (a) polar  and (b) antipolar  structures. There are four structural domains in each orthorhombic twin, which are indicated by the  red and blue shaded regions. For each structural domain, the OPs $\eta^{X_3^-}$ and $\eta^{X_2^+}$ are shown (red and green arrows), as well as the corresponding structural distortions. Both $X_3^-\oplus X_2^+$ domains in each quadrant trilinearly couple to  the same domain of ${\bf \eta}^P$ or ${\bf \eta}^{AP}$ (blue and purple arrows).  (c) shows an illustration of a ``stacking domain wall" (dashed line) between perovskite slabs  with polarization ${\bf P}$ and slabs with polarization $-{\bf P}$. The crystallographic structure at this stacking domain wall is the antipolar structure.  The blue arrows indicate the polarization direction in each perovskite slab.}
 \end{center}
 \end{figure*}

The two-dimensional OPs that describe the distortions that transform like $X_3^-$ and $X_2^+$ are:~\cite{harris11,stokes1988}
\begin{eqnarray}
{\mathbf \eta}^{X_3^-} & = & (\eta^{X_3^-}_1,\eta^{X_3^-}_2)  = Q_{X_3^-}e^{i\theta_{X_3}}\nonumber \\
{\bf \eta}^{X_2^+} & = & (\eta^{X_2^+}_1,\eta^{X_2^+}_2) = Q_{X_2^+}e^{i\theta_{X_2}}.
\label{i4mmm_ops}
\end{eqnarray}  
Taking the example of $X_3^-$, 
there are three symmetry-inequivalent choices  for the OP (directions of the irrep): $(\eta^{X_3^-}_1,\eta^{X_3^-}_2)$ = $(a,0)$, $(a,a)$, and $(a,b)$, where $a$ and $b$ are real numbers. Each choice defines one of the three isotropy  subgroups of $I4/mmm$ generated by $X_3^-$, and thus describes a structural distortion of different symmetry. 
These isotropy subgroups are listed in Table~\ref{X3X2}, as well as those generated by $X_2^+$. 

The OPs $(a,0)$ and $(0,a)$ define orthorhombic twin domains of the same isotropy subgroup.~\cite{stokes1988}
The orientation of the orthorhombic relative to the tetragonal axes,  shown  in Fig.~\ref{struc}(d), determines the setting of the space group symbol  that describes the symmetry of each twin domain. 
The  $(a,0)$ and $(0,a)$ directions of $X_3^-$ define twin domains  $Amam$ and $Bbmm$, the structural distortions in these domains, shown in  Fig.~\ref{struc}(e, left column), are out-of-phase  octahedral tilts  about [110] and [-110], respectively.
Similarly, for $X_2^+$, the $(a,0)$ and $(0,a)$ directions define twin domains  $Acam$ and $Bbcm$,  in both domains there is  an  in-phase octahedral rotation about [001], but they have different relative rotation senses in adjacent  perovskite slabs (Fig.~\ref{struc}(e, right column)).

Table~\ref{X3X2} also lists the isotropy subgroups generated by the coupled irreps $X_3^-\oplus X_2^+$. 
The two orthorhombic twins of the polar ground state have symmetry $A2_1am$ and $Bb2_1m$. The combination of  OPs $\{\eta^{X_3^-}=(a,0)$, $\eta^{X_2^+}=(b,0)\}$  establishes $A2_1am$, while $\{\eta^{X_3^-}=(0,a)$, $\eta^{X_2^+}=(0,b)\}$ establishes $Bb2_1m$, as illustrated in Fig.~\ref{struc}(e) (top and bottom).
Within each twin domain, there are four structural domains. To generate the other structural domains from the representative ones shown in Fig.~\ref{struc}(e), consider all possible ways of combining the   
 two octahedral rotation distortions, each of which has two senses. This leads to a total of eight structural domains, shown in Fig.~\ref{domains}(a). 
By using OPs restricted to one dimension, Ref.~\onlinecite{benedek11} considered only one twin domain,  so the four domains  discussed in that work are those of $A2_1am$.

The same directions of OPs $\eta^{X_3^-}$ and $\eta^{X_2^+}$, but  taken in different combinations, define 
the two orthorhombic twin domains of the antipolar structure, which have symmetry  $Pnam$ and $Pbnm$.  $Pnam$  is established by   $\{\eta^{X_3^-}=(a,0)$, $\eta^{X_2^+}=(0,b)\}$, while  $Pbnm$ is established by $\{\eta^{X_3^-}=(0,a)$, $\eta^{X_2^+}=(b,0)\}$, as shown in Fig.~\ref{struc}(f) (top and bottom).  
Comparing polar $A2_1am$ to antipolar $Pnam$ (Fig.~\ref{struc}(e,f), top row),  one can see that the only  difference between the octahedral rotation distortions in these two structures is the sense of the in-phase octahedral rotation about [001] in perovskite slab 2.
Therefore, it is unsurprising that we find antipolar $Pnam$ to be only 7 meV/Ti above the polar ground state. 
Each antipolar twin domain has four structural domains, leading to a total of eight structural domains,  shown in Fig.~\ref{domains}(b). 
The other $X_3^- \oplus X_2^+$ isotropy subgroups listed in Table~\ref{X3X2} are 30-40 meV/Ti  above the polar ground state, and  will be important for our discussion of ferroelectric switching in the next section.  The domains of these subgroups, and the   structural properties of \cto~ in all subgroups, are given in Appendix~\ref{domains-app}.

There are additional trilinear terms when the Landau expansion of the energy is performed with two-dimensional OPs.
Whenever bilinears  $\eta^{X_3^-}_\alpha\eta^{X_2^+}_\alpha$ and  $\eta^{X_3^-}_\alpha\eta^{X_2^+}_\beta$ ($\alpha,\beta=1,2$, $\alpha \ne \beta$) have the same transformation properties under the generators of $I4/mmm$ as a direction of an irrep of a different symmetry, a trilinear coupling is allowed (see Appendix~\ref{tri-app} for the transformation properties).  We find that the allowed trilinear terms are:~\cite{foot1}
\begin{eqnarray}
\mathcal{F}_{\mathrm{tri}} &=& \alpha\eta^{X_3^-}_1 \eta^{X_2^+}_1 \eta^{P}_2 + \beta\eta^{X_3^-}_2 \eta^{X_2^+}_2 \eta^{P}_1 \nonumber \\
&&+ \gamma\eta^{X_3^-}_1 \eta^{X_2^+}_2 \eta^{AP}_2 + \delta\eta^{X_3^-}_2 \eta^{X_2^+}_1 \eta^{AP}_1.
\label{eq:tri-2d}
\end{eqnarray}
Here $\eta^P=Q_Pe^{i\theta_P}=(\eta_1^P,\eta_2^P)$ is the OP for the polar distortion that transforms like \gfive, and $\eta^{AP}=Q_{AP}e^{i\theta_{AP}}=(\eta_1^{AP},\eta_2^{AP})$ is the OP for the antipolar distortion that transforms like irrep $M_5^-$ (both these irreps and hence their OPs are two-dimensional). 
The $(a,0)$ direction of $\eta^P$ ($\eta^{AP}$) describes a polar (antipolar) distortion with displacements  along [-110], while $(0,a)$ describes a polar (antipolar) distortion with displacements along [110].
 
 In  polar twin domain $A2_1am$, the first term in Eq.~\ref{eq:tri-2d} is nonzero (this is the term given in Eq.~\ref{eq:tri-1d}), and all  other terms vanish, while in polar twin domain $Bb2_1m$, only the second term in Eq.~\ref{eq:tri-2d} survives.  In the two twin domains of the antipolar structure, $Pnam$ and $Pbnm$, only the third and fourth terms  survive, respectively. For all other $X_3^-\oplus X_2^+$ istotropy subgroups, there is more than one trilinear coupling term.  For example, all four terms in  Eq.~\ref{eq:tri-2d} are nonzero  in $C2mm$, so both components of $\eta^P$ and $\eta^{AP}$ are induced, see Table~\ref{X3X2}. 
In  $P2_1am$ and $P2_1nm$ (and their twins), there are two nonzero trilinear coupling terms (one to a component of $\eta^P$, one to a component of $\eta^{AP}$). 
  
To summarize, we have found  that the antipolar structure is only slightly higher energy than the polar ground state, and that the domains of these two structures are compatible. This has an important implication: all eight polar domains and all eight antipolar domains shown in Fig.~\ref{domains} should be present in a bulk as-grown multidomain sample. The eight polar domains, and a complex network of  domain walls between them, have already  been detected experimentally.~\cite{huang2016, huang2016-2}
An additional type of  domain wall that should be present in samples is what we call a ``stacking domain wall," illustrated in Fig.~\ref{domains}(c): an interface between a polar domain  with polarization ${\bf P}$ stacked on top of (along [001]) a polar domain with polarization $-{\bf P}$.  At this stacking domain wall, which lies parallel to the $ab$ plane, the structure locally is  antipolar, so if there are stacking domain walls in the sample,  antipolar domains are necessarily present. While our bulk calculation found antipolar $Pnam$ to be 7 meV/Ti above the polar ground state, the energy to form a stacking domain wall between many-unit cell polar domains  should become vanishingly small.
We suggest atomic scale imaging of the local Ca displacements to observe these stacking domain walls experimentally.  
The appropriate starting point for considering possible ferroelectric switching paths in the next section   is to treat all polar and antipolar domains, as well as the twin domain walls and stacking domains walls between them, on an equal footing.

\section{Ferroelectric switching paths\label{section-switching}}
\begin{figure}
 \includegraphics[width=0.48\textwidth]{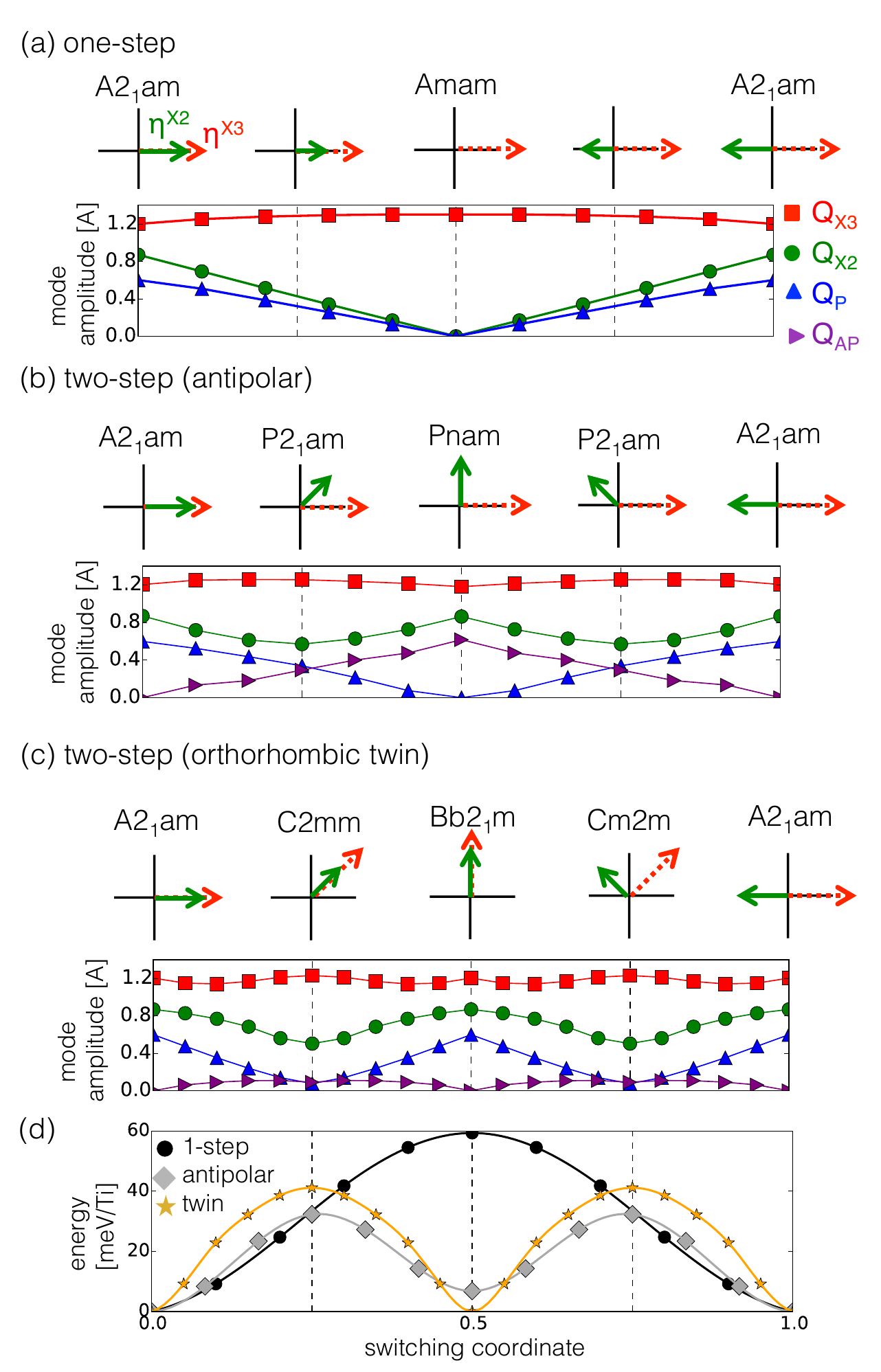}
 \caption{\label{fig2} Ferroelectric  switching pathways that reverse the $X_2^+$ octahedral rotation. (a) one-step switching, (b, c) two-step switching via antipolar ($Pnam$) and orthorhombic twin ($Bb2_1m$) intermediate, respectively. The top of each panel shows the $\eta^{X_3^-}$ (red) and $\eta^{X_2^+}$ (green) OPs, while the bottom shows the amplitudes of all OPs, obtained by decomposing the structures into symmetry adapted modes.~\cite{perez-mato10} (d) The total energy as a function of switching coordinate for paths shown in (a-c). }
 \end{figure}

In this section we enumerate the possible ferroelectric switching paths in twin domain $A2_1am$ (paths in twin $Bb2_1m$ are analogous). 
As discussed in the Introduction, reversing the polarization direction ($\eta^P\rightarrow -\eta^P$) requires reversing one but not both of the octahedral rotation OPs $\eta^{X_3^-}$ and $\eta^{X_2^+}$. Here we  discuss switching paths that reverse $\eta^{X_2^+}\rightarrow -\eta^{X_2^+}$,  those that reverse $\eta^{X_3^-}$ are shown in Appendix~\ref{X3-app}.  We identify three distinct  types of paths that reverse $\eta^{X_2^+}$, which are presented in Fig.~\ref{fig2}. 

In the first path, shown in Fig.~\ref{fig2}(a), $\eta^{X_2^+}$ reverses by turning off  and then turning back on, pointing in the opposite direction. At the energy barrier ($Amam$), where $Q_{X_2^+}=0$, the polarization is also zero, because all trilinear coupling terms in Eq.~\ref{eq:tri-2d} vanish.  We refer to this as a one-step switching path; it previously was discussed in Ref.~\onlinecite{benedek11} (one-dimensional OPs are sufficient to describe this path). 

In the second path, shown in Fig.~\ref{fig2}(b), $\eta^{X_2^+}$ reverses by rotating in OP space (changing its phase $\theta_{X_2}$ by $\pi$, while its amplitude $Q_{X_2^+}$ stays finite). When $\theta_{X_2}=\pi/2$, the crystallographic structure is antipolar $Pnam$, due to the presence of this low-energy intermediate state, we refer to this as a two-step switching path. Both OP amplitudes $Q_{X_3^-}$ and $Q_{X_2^+}$ are nonzero throughout the switching process, although $Q_{X_2^+}$ decreases at the energy barrier ($P2_1am$).~\cite{foot7}  As a result, there are nonzero trilinear couplings, which induce $\eta^P$ and/or $\eta^{AP}$,  throughout the switching process (specifically, because $\eta^{X_3^-}=Q_{X_3^-}(1,0)$ stays fixed, the relevant trilinear couplings are the first and third terms of Eq.~\ref{eq:tri-2d};  as $\eta^{X_2^+}$ rotates,  the magnitudes of    $\eta_1^{X_2^+}$  and $\eta_2^{X_2^+}$, and therefore these two trilinear terms, change).

Finally, Fig.~\ref{fig2}(c) shows the third type of path, here $\eta^{X_2^+}$ reverses by rotating in OP space as in Fig.~\ref{fig2}(b), but now $\eta^{X_3^-}$ also rotates. When $\theta_{X_2}=\theta_{X_3}=\pi/2$, the crystallographic structure is the polar orthorhombic twin $Bb2_1m$; this is also a two-step switching process. As in Fig.~\ref{fig2}(b), both $Q_{X_3^-}$ and $Q_{X_2^+}$ are nonzero throughout the switching process, with $Q_{X_2^+}$ decreased by about half at the energy barrier ($C2mm$),~\cite{foot8} so there are nonzero trilinear couplings inducing $\eta^P$ and/or $\eta^{AP}$ throughout. In this case, because both  $\eta^{X_3^-}$ and $\eta^{X_2^+}$ rotate, all four terms in Eq.~\ref{eq:tri-2d} are active. The polar OP $\eta^P$ is nonzero throughout switching, although its amplitude decreases significantly at the energy barrier.
Fig.~\ref{fig2}(d) shows the total energy as a function of switching coordinate, we find that the two-step paths both have lower energy barriers  than the one-step path, with the antipolar path barrier being the lowest. 
 
Table~\ref{switching} summarizes the properties of the switching paths presented in Fig.~\ref{fig2}, as well as those that reverse $\eta^{X_3^-}$. These are all  the possible switching paths, of two steps or less, within the space defined by $X_3^-\oplus X_2^+$. 
We reach a different understanding of the  possible switching processes by using the full two-dimensional OPs (the one-dimensional OPs only can describe the one-step paths).
 First, the predicted energy barriers are significantly lower: the barrier for switching via reversing $\eta^{X_2^+}$ reduces from 56 meV/Ti (one-step)  to 32 meV/Ti  (two-step), while for switching via $\eta^{X_3^-}$ reversal, it reduces from 90 to 35 meV/Ti.
 Second, there is a large  difference in barrier energy between the one-step path that reverses $\eta^{X_3^-}$, and the one-step path that reverses $\eta^{X_2^+}$ (90 versus 56 meV/Ti). However, the barrier for two-step switching via the antipolar intermediate ($Pnam$ or $Pbnm$) is almost the same for reversing $\eta^{X_3^-}$ or $\eta^{X_2^+}$ (35 versus 32 meV/Ti), while the barrier for switching via  orthorhombic twin $Bb2_1m$ is the same regardless of which octahedral rotation is reversed (compare Fig.~\ref{fig2}(c) and Appendix~\ref{X3-app}, Fig.~\ref{x3-switching}(c)).
These energetics suggest that ferroelectric switching may proceed via more than one mechanism in \cto.

Two-step ferroelectric switching paths  have been discussed previously in a couple other systems.
 A  first principles study of  AFeO$_3$/A$^\prime$FeO$_3$ superlattices, which are predicted to be hybrid improper ferroelectrics,  found a two-step switching path to have the lowest energy barrier.~\cite{zanolli13} 
 In addition, it has been experimentally demonstrated, and corroborated by first principles calculations, that  ferroelectric switching in BiFeO$_3$  follows a two-step path~\cite{heron14}  (although the details are different, due to the different crystallographic structures and ferroelectric mechanisms in BiFeO$_3$ and  \cto). 

\begin{table}[]

\caption{\label{switching}Summary of switching paths. }

\begin{tabular}{c | c | c |   c  | c}
\hline
\hline
path & barrier & intermediate & E$_{\mathrm{barrier}}$& OP  \\
& & & [meV/Ti] & reversed \\
\hline
one-step & $Amam$ & -& 56 & $\eta^{X_2^+}$ \\
one-step & $Acam$ & - & 90 & $\eta^{X_3^-}$ \\
two-step, AP  & $P2_1am$ & $Pnam$ & 32 & $\eta^{X_2^+}$ \\
two-step, AP & $P2_1nm$ & $Pbnm$ & 35 & $\eta^{X_3^-}$ \\
two-step, twin & $C2mm$ & $Bb2_1m$ & 41 & $\eta^{X_2^+}$ or $\eta^{X_3^-}$  \\
\hline
\end{tabular}
\end{table}

\section{Local order parameters to elucidate switching paths\label{section-local}}
How do the  two-step switching paths  identified in the previous section relate to the crystallographic structure and switching in the real material, which is likely nucleated at domain walls?
To gain insight into these questions, we introduce a set of OPs local to each $n=2$ perovskite slab in \cto. 
This will allow us to rewrite Eq.~\ref{eq:tri-2d} in terms of trilinear couplings local to each slab, and clearly visualize how the crystallographic structure evolves during the switching process.    
To do this, we write each structural distortion ${\bf u}^Y$ that transforms like irrep $Y=\{X_3^-,X_2^+, P, AP\}$ as a sum of distortions local to the two adjacent perovskite slabs $\alpha$ in the unit cell: ${\bf u}^Y = \sum_{\alpha=1,2}{\bf u}^{(\alpha)}$, and introduce an OP local to each slab to describe their symmetry properties. These local distortions, written in Glazer notation,~\cite{glazer72} are an $a^-a^-b^0$ tilt, an $a^0a^0b^+$ rotation, and a polar distortion in each perovskite slab (for the polar and antipolar structures, for arbitrary irrep directions the rotation pattern is $a^-b^-c^+$).
A similar local OP approach was used in Ref.~\onlinecite{hena} to describe the hexagonal manganites. 

We define each of these local structural distortions and the OPs that describe them in Fig.~\ref{fig3} (see Appendices~\ref{tri-app} and~\ref{local-app} for complete details). 
The $X_3^-$ distortion (with arbitrary irrep direction) can be written as the sum of  $a^-b^-c^0$ tilt distortions  ${\bf u}^{(\alpha)}$  in slabs $\alpha=1,2$.  We introduce the local two-dimensional OP $T_\alpha = (T_{\alpha x},T_{\alpha y})=Q_{T\alpha} e^{i\theta_\alpha}$ to describe ${\bf u}^{(\alpha)}$, $T_\alpha = Q_{T\alpha}(1,0)$ and $Q_{T\alpha}(0,1)$ describe  $a^-a^-b^0$ tilts about [110] and [-110] in slab $\alpha$, respectively (see Fig.~\ref{fig3}(a,b)).
The phase $\theta_\alpha$ can be interpreted as the angle of the apical oxygen displacement vector in slab $\alpha$. The symmetry  of the $X_3^-$ distortion imposes a relation between the $T_\alpha$ in adjacent slabs: $Q_{T1} = Q_{T2}=Q_T$, and $\theta_1 = -\theta_2$. 
Similarly, we express the $X_2^+$ distortion as the sum of $a^0a^0b^+$ rotation distortions in  slabs $\alpha=1,2$ (see Fig.~\ref{fig3}(c,d)), and introduce the local OP $\Phi_\alpha$ to describe them. The local OP $\Phi_\alpha$ is one dimensional because the rotation axis is fixed to [001]. 

Both the polar and antipolar distortions (along arbitrary direction) can be written as the sum of local polar distortions in slabs $\alpha=1,2$, which we describe with the local two-dimensional OP $P_\alpha=(P_{\alpha x},P_{\alpha y})=Q_{P\alpha} e^{i\phi_\alpha}$. Here  $P_\alpha = Q_{P\alpha}(1,0)$ and $Q_{P\alpha}(0,1)$ describe  polar displacements along [-110] and [110] in slab $\alpha$, respectively (Fig.~\ref{fig3}(e,f)).
 The phase $\phi_\alpha$ can be interpreted as the angle of the polarization vector in slab $\alpha$.
 
The transformation properties of these local OPs, as well as the bilinear combination $Y_\alpha = P_{\alpha x}T_{\alpha y}-P_{\alpha y}T_{\alpha x}$, under the generators of $I4/mmm$ are shown in Appendix~\ref{tri-app}.  We find the invariant trilinear terms, and  reexpress Eq.~\ref{eq:tri-2d} in terms of the local OPs:
\begin{equation}
\mathcal{F}_\mathrm{tri} = \beta_1Y_1\Phi_1 + \beta_2Y_2 \Phi_2.
\label{eq:local_tri_1}
\end{equation}
Therefore,  the  four trilinear coupling terms in Eq.~\ref{eq:tri-2d} reduce to just two trilinear terms, each of which is local to one  perovskite slab.

We can bring Eq.~\ref{eq:local_tri_1} to a more illuminating form by representing  the local OPs as vectors (note that we are free to choose to represent each octahedral rotation as either a polar or an axial vector).~\cite{foot4} Writing $T_\alpha$ and $P_\alpha$ as polar vectors (${\bf T}_\alpha = Q_{T\alpha}(\cos \theta_\alpha \hat{\bf x} + \sin\theta_\alpha\hat{\bf y})$, ${\bf P}_\alpha = Q_{P\alpha}(\cos \phi_\alpha \hat{\bf x} + \sin\phi_\alpha\hat{\bf y})$), and $\Phi_\alpha$ as an axial vector (${\bf \Phi}_\alpha = \Phi_\alpha \hat{\bf z}$), Eq.~\ref{eq:local_tri_1} becomes:
\begin{equation}
\mathcal{F}_{\mathrm{tri}} =\beta_1({\bf T}_1 \times {\bf \Phi}_1) \cdot {\bf P}_1 + \beta_2({\bf T}_2 \times {\bf \Phi}_2) \cdot {\bf P}_2.
\label{eq:local_tri}
\end{equation}
Therefore, the two octahedral rotation OPs in slab $\alpha$, via a right-hand rule, determine the polarization direction in slab $\alpha$;  reversing   one of these octahedral rotation OPs  reverses the polarization direction in that slab (taking the structure from being polar to antipolar, or vice versa).

\begin{figure}
 \includegraphics[width=0.48\textwidth]{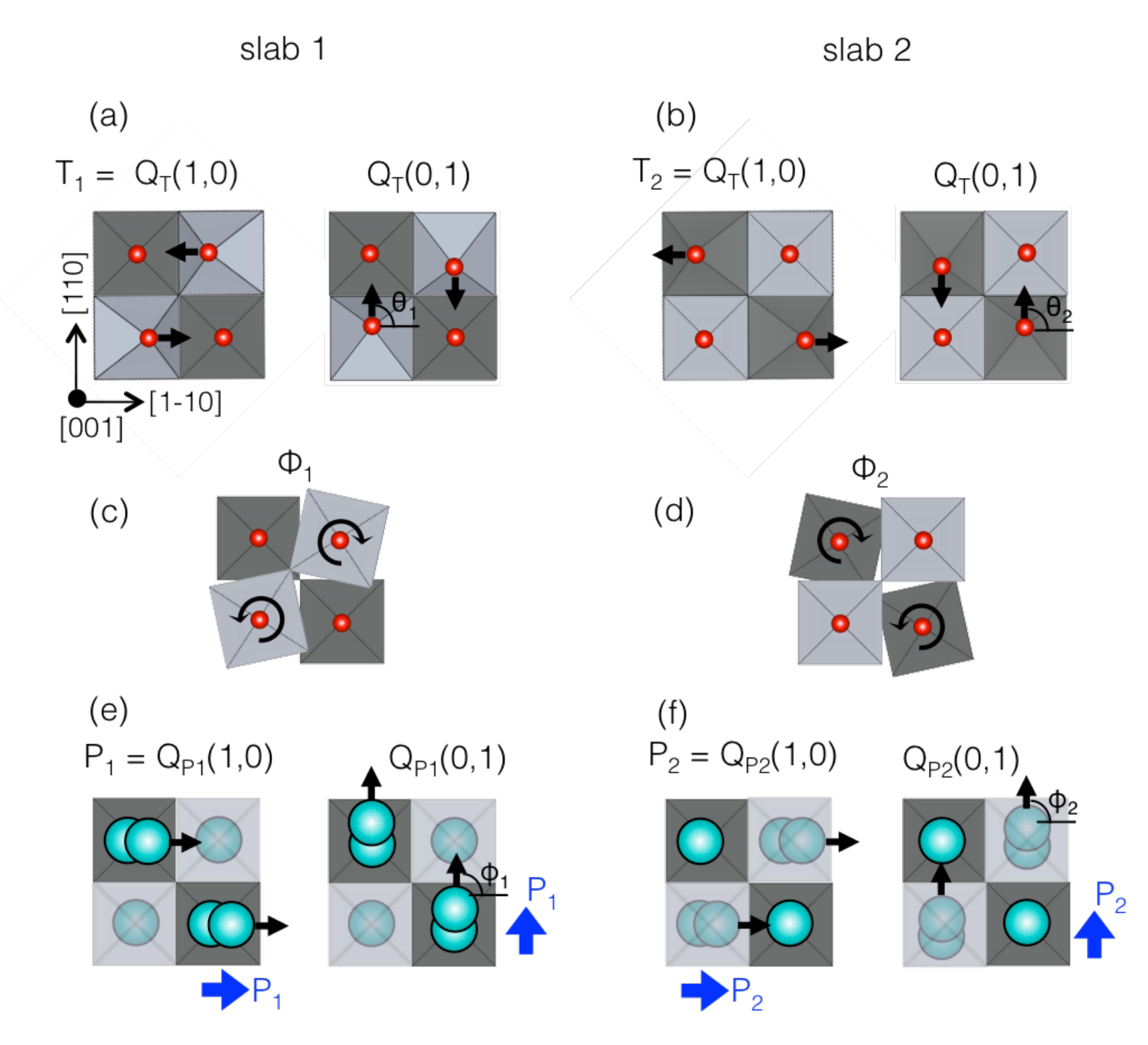}
 \caption{\label{fig3} Definition of local order parameters (OPs). Structural distortions local to perovskite slab 1 (2) are shown in the left (right) column, each of the local distortions is described by an OP local to that slab. (a, b) show out-of phase $a^-a^-b^0$ tilt distortions described by two-dimensional OP $T_\alpha$ in slab $\alpha$, (c, d) show the in-phase $a^0a^0b^+$ rotation distortions described by OP $\Phi_\alpha$, and (e, f) show the polar distortions described by two-dimensional OP $P_\alpha$. }
  \end{figure} 

\begin{figure}
 \includegraphics[width=0.48\textwidth]{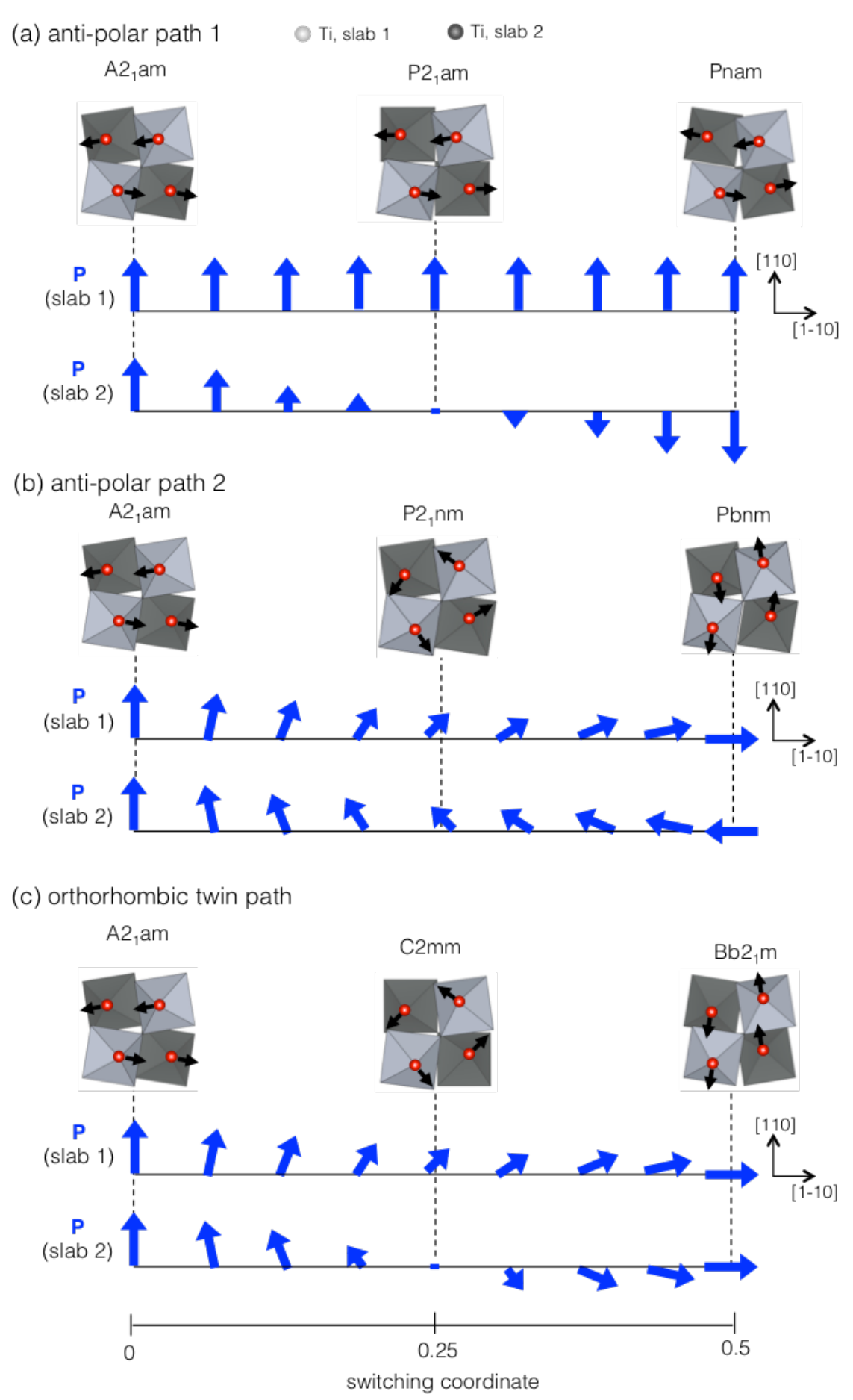}
 \caption{\label{fig4} Crystallographic structure and polarization in perovskite slabs 1 and 2 during switching  from polar  $A2_1am$ to  (a) antipolar $Pnam$, (b)  antipolar orthorhombic twin $Pbnm$, and (c)  polar orthorhombic twin $Bb2_1m$. The top of each panel shows the crystallographic structure at three points along the switching path (switching coordinate = 0, 0.25, 0.5), note that 0.25 is the barrier structure. The black arrows indicate the apical oxygen motion under the combined tilt and rotation distortions in each slab.  The bottom part of each panel shows the polarization in slabs 1 and 2 as a function of switching coordinate.}
 \end{figure}

With these local OPs, we return to the question of how the crystallographic structure changes during each of the two-step switching paths listed in Table~\ref{switching}. Fig.~\ref{fig4} shows the first step of each path (the second step is similar). 
Fig.~\ref{fig4}(a) shows switching from polar $A2_1am$ to antipolar $Pnam$, along this path, the polarization in slab 1, ${\bf P}_1$, stays  fixed, while the polarization in slab 2, ${\bf P}_2$, reverses. The only difference between $A2_1am$ and $Pnam$ is the $a^0a^0b^+$ rotation sense in slab 2 (${\bf \Phi}_2^{A2_1am} =-{\bf \Phi}_2^{Pnam}$), so at the energy barrier ${\bf \Phi}_2=0$, by symmetry. This barrier structure, $P2_1am$, has rotation pattern $a^-a^-b^+$ in slab 1 and $a^-a^-b^0$ in slab 2, so only ${\bf P}_1$ is induced (via the first term in Eq.~\ref{eq:local_tri}), while ${\bf P}_2=0$.

The second antipolar switching path ($A2_1am\rightarrow Pbnm$) is shown in Fig.~\ref{fig4}(b), in this case the polarization vectors in both slabs rotate. The difference between $A2_1am$ and $Pbnm$ is the axis of the $a^-a^-b^0$ tilt, this axis rotates by $-(+)\pi/2$ in slab 1(2) when switching from $A2_1am\rightarrow Pbnm$. By symmetry, at the energy barrier ($P2_1nm$), slab 1 has rotation pattern $a^-b^0c^+$ with polarization along [100], while slab 2 has rotation pattern $b^0a^-c^+$ with polarization along [010] (both terms in Eq.~\ref{eq:local_tri} are nonzero). Note that while ${\bf P}_1$ and ${\bf P}_2$ rotate in opposite directions, $|{\bf P}_1|=|{\bf P}_2|$ throughout, so the net polarization stays fixed along [110], with amplitude diminishing to zero at antipolar $Pbnm$.

Fig.~\ref{fig4}(c) shows switching between orthorhombic twins $A2_1am$ and $Bb2_1m$. To go between these two structures, the $a^0a^0b^+$ rotation sense in slab 2 must reverse, and the $a^-a^-b^0$ tilt axes must rotate, so one can loosely think of this path as the ``sum" of the two antipolar paths in Fig.~\ref{fig4}(a,b).  At the energy barrier ($C2mm$), by symmetry slab 1 has rotation pattern $a^-b^0c^+$ with polarization along [100], while slab 2 has rotation pattern $b^0a^-c^0$ with ${\bf P}_2=0$. Along this path,  the total polarization rotates from [110] to [-110]. 

In both the  $P2_1nm$ and  $C2mm$ barrier structures (Fig.~\ref{fig4}(b,c)), the tilt axes in adjacent perovskite slabs are  perpendicular to each other (tilt pattern $a^-b^0c^0$ in one slab and $b^0a^-c^0$ in the other). Interestingly, considering isotropy subgroups generated by   $X_3^-$ only, we find that this tilt pattern ($\eta^{X_3^-} = (a,a)$,  $P4_2/mnm$) is in fact lower  energy than the $a^-a^-b^0$ tilt pattern  in the polar ground state ($\eta^{X_3^-}=(a,0)$, $Amam$), see Table~\ref{X3X2}. 
Therefore, the  coupling to the $X_2^+$ and polar distortions stabilizes the $\eta^{X_3^-}=(a,0)$ tilt pattern in $A2_1am$. By changing the energetics of these different distortions using chemical doping or epitaxial strain, one may be able to stabilize the tetragonal $P4_2/mnm$ structure.  
It was recently found that Ca$_{3-x}$Sr$_x$Ti$_2$O$_7$ with $x\sim 1$ crystallizes in space group $P4_2/mnm$,~\cite{huang2016-2} as does SrTb$_2$Fe$_2$O$_7$.~\cite{pitcher15}

To summarize, all three of these two-step paths take advantage of  the fact that the perovskite slabs are disconnected across the rocksalt layer, which allows the polarizations local to each perovskite slab, via the local trilinear couplings in Eq.~\ref{eq:local_tri}, to turn off and/or rotate independently.

\section{Discussion\label{section-discussion}}
We now discuss how the two-step switching paths relate to twin and stacking domain walls.
 The orthorhombic twin path (Fig.~\ref{fig2}(c), \ref{fig4}(c), \ref{x3-switching}(c)) describes switching at a twin domain wall, as was previously suggested in Ref.~\onlinecite{oh15}. Our finding  that this path's intrinsic energy barrier is the same regardless of whether $\eta^{X_3^-}$ or $\eta^{X_2^+}$ reverses suggests that both variations of this path could contribute to the switching process. 
 This may help explain the recent observation of multiple types of twin domain walls in Ca$_3$Ti$_2$O$_7$, where $\eta^{X_3^-}$,  or both $\eta^{X_3^-}$ and $\eta^{X_2^+}$,  change across the wall.~\cite{huang2016}
The antipolar path, shown  in Fig.~\ref{fig2}(a), \ref{fig4}(a), describes switching at the stacking domain wall introduced in Fig.~\ref{domains}(c). That is, as the polarization reverses in individual perovskite slabs stacked along [001], this stacking domain wall would move up or down.   
Finally, the antipolar path shown in Fig.~\ref{fig4}(b),\ref{x3-switching}(b) describes switching at a twin wall between a polar and an antipolar domain.
Depending on the prevalence of twin and stacking domain walls in a given material, and the relative energetics to move these walls, any one of these paths could be the dominant switching mechanism. 

Finally, by focusing on switching paths that stay in the space of $X_3^- \oplus X_2^+$ isotropy subgroups, we are restricted to structures with  $a^-a^-b^+$ rotations. What about switching via a structure with a different rotation pattern?  Using intuition from perovskites,~\cite{woodward97b}  structures with $a^-a^-b^-$ rotations also may  be low energy. We identify two such structures, $Pbna$ and $C2/c$, established by $X_3^-\oplus X_1^-$, and find that their energies are 14 and 21 meV/Ti above $A2_1am$, respectively.~\cite{foot6} Our computed energy barriers for two-step polarization switching with $Pbna$ ($C2/c$) as an intermediate are 30 (34) meV/Ti, approximately the same  as the antipolar switching barriers. While it is not immediately clear how switching via these rhombohedral phases could be nucleated, this may be an interesting topic for future investigation (a recent first principles study found that epitaxial strain stabilizes $Pbna$~\cite{lu2016}).

In summary, this work  uses  two-dimensional OPs to describe the polar and antipolar domains of Ca$_3$Ti$_2$O$_7$, and enumerates possible ferroelectric switching pathways. We find switching via an orthorhombic twin domain,  an antipolar stacking domain, or a rhombohedral-like phase, to have low energy barriers. 
In addition, we find that the antipolar structure $Pnam$  is only slightly higher energy than the polar ground state. An interesting direction for possible future investigations is understanding how to control, via chemical doping, the relative energetics of these two structures. 
This could help elucidate the relationship between ferroelectricity and antiferroelectricity, and 
 provide a pathway to find new antiferroelectrics, which would be of both fundamental and technological interest. 

The authors thank D. G. Schlom for useful discussions. We acknowledge support from the Army Research Office under grant number W911NF-10-1-0345.

\appendix

\section{\label{domains-app}Domains and structural properties of $X_3^-\oplus X_2^+$ isotropy subgroups}

\begin{figure*}
\begin{center}
 \includegraphics[width=0.65\textwidth]{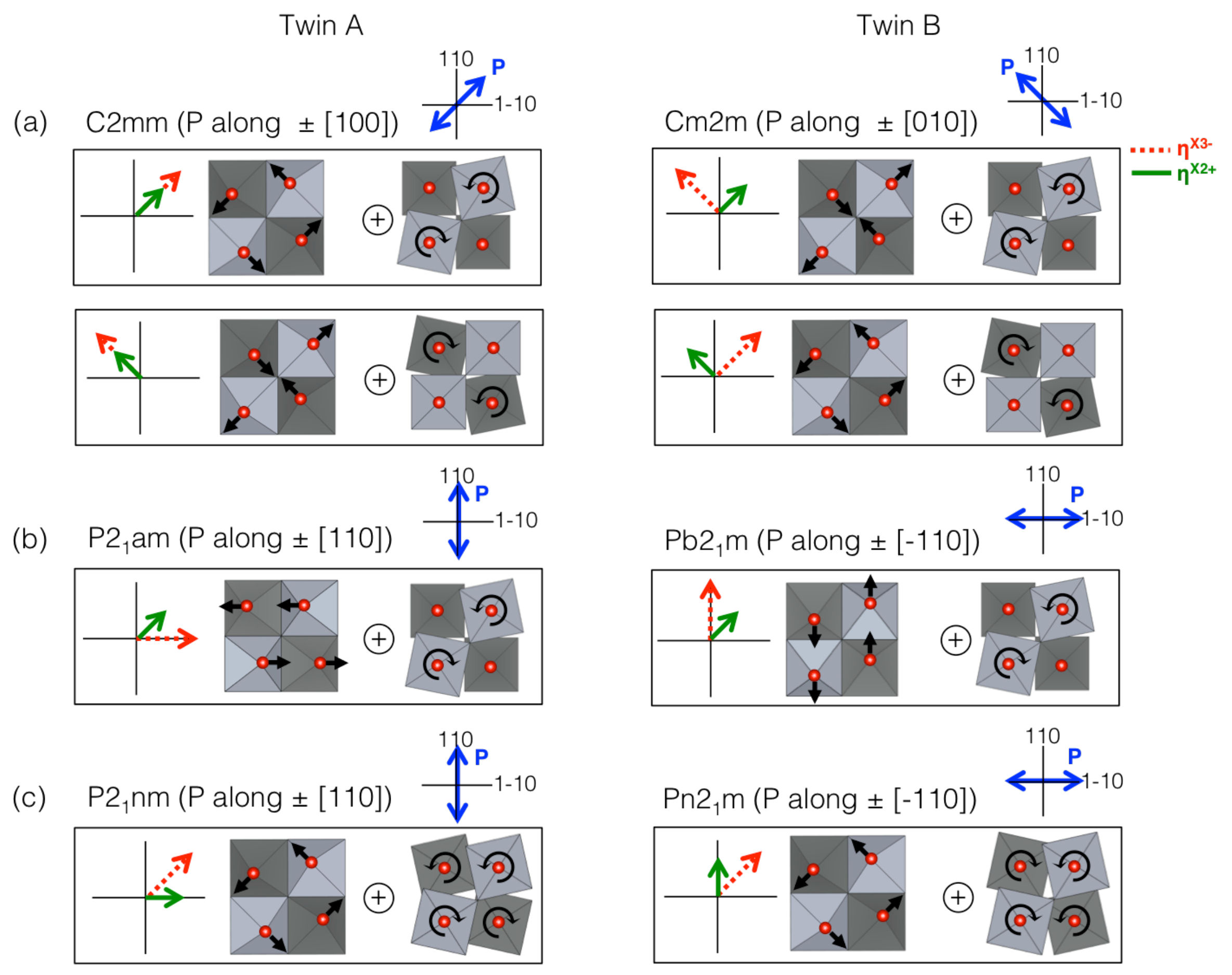}
 \caption{\label{more-domains}  Distinct domains of $X_3^-\oplus X_2^+$ isotropy subgroups not shown in the main text. The two twin domains are shown in the left (right) column: (a) $C2mm$ ($Cm2m$), (b) $P2_1am$ ($Pb2_1m$), and (c) $P2_1nm$ ($Pn2_1m$).  For each distinct domain, the OPs $\eta^{X_3^-}$ and $\eta^{X_2^+}$ are shown, as well as the structural distortions described by these OPs. The polarization directions in each twin domain are shown by blue arrows. In (b) $\eta^{X_3^-}=(a,b)$ and in (c) $\eta^{X_2^+}=(a,b)$ are along general directions, here we show the structures when $a=b$.}
 \end{center}
 \end{figure*}

Fig.~\ref{more-domains} shows the distinct domains of the isotropy subgroups $C2mm$ (twin $Cm2m$), $P2_1am$ (twin $Pb2_1m$), and $P2_1nm$ (twin $Pn2_1m$). Each of these subgroups has 16 total structural domains. As shown in Fig.~\ref{more-domains}(a),   $C2mm$ ($Cm2m$) has four distinct domains  established by the combination of OPs $\{\eta^{X_3^-} = (a,a), \eta^{X_2^+} = (b,b)\}$, $\{\eta^{X_3^-} = (-a,a), \eta^{X_2^+} = (-b,b)\}$, $\{\eta^{X_3^-} = (a,a),\eta^{X_2^+} = (-b,b)\}$,  and $\{\eta^{X_3^-} = (-a,a), \eta^{X_2^+} = (b,b)\}$. In the first two domains, the polarization lies along [100], while in the second two it is along [010]. All 16 domains can be generated by considering all ways to combine the two senses of the $X_3^-$ and $X_2^+$ octahedral rotations shown in the four distinct domains.   The other subgroups $P2_1am$ ($Pb2_1m$) and $P2_1nm$ ($Pn2_1m$) have two distinct domains, shown in Fig.~\ref{more-domains}(b-c), in these domains the polarization lies along [110] ([-110]).  All 16 domains can be obtained from the two distinct ones shown by considering all ways to combine the two senses of the high-symmetry direction OP $(\pm a,0)$ and the four possibilities for the general direction OP $(b,c)$, $(-b,c)$, $(b,-c)$ and $(-b,-c)$.

Table~\ref{mystructures} lists the structural parameters of \cto~ in all isotropy subgroups.

\begin{table*}[]
\begin{center}
\caption{\label{mystructures} Structural parameters of \cto~  in all isotropy subgroups where both OPs are along high-symmetry directions. The OP amplitudes are obtained by decomposing the structures into symmetry adapted modes,~\cite{perez-mato10} and are given for a $Z$=2 cell. 
}

\begin{tabular}{c | c | c | c | c | c | c | c | c }
\hline
\hline
 & $I4/mmm$ & $Amam$ & $P4_2/mnm$ & $Acam$ & $P4/mbm$ & $A2_1am$ & $Pnam$ & $C2mm$ \\ 
 \hline
 $a$ [\AA] & 5.42 & 5.39 &      5.45 &                5.33 &        5.38 &         5.39 &             5.39 &          7.67 \\ 
 $b$ [\AA] & 5.42 & 5.48 &      5.45    &               5.33 &       5.38 &        5.44  &             5.43 &          7.67  \\ 
 $c$ [\AA] & 19.40 & 19.14  &  19.05    &            19.88 &     19.61 &       19.30   &         19.36 &        19.17  \\ 
\hline
$Q_{X_3^-}$ [\AA] & - & 1.3 &  1.87 & - &          - &                 1.20 & 1.18 & 1.74  \\ 
$Q_{X_2^+}$  [\AA] & - & -          &       -         &  1.21 & 1.09 & 0.87 & 0.86 & 0.71  \\ 
$Q_{P}$  [\AA] & - & -          &   -         & -          & - &                  0.60 &     -       & 0.11  \\ 
$Q_{AP}$  [\AA] & - & -          &            -           &  -           & - &                  - &           0.62   & 0.13  \\ 
\hline 
P $[\mu C/cm^2]$ &0 & 0& 0& 0& 0& 17 & 0 & 0.9  \\ 
\hline
\end{tabular}
\end{center}
\end{table*}


\section{\label{X3-app}Switching paths that reverse $X_3^-$ octahedral rotation sense}
\begin{figure}
 \includegraphics[width=0.45\textwidth]{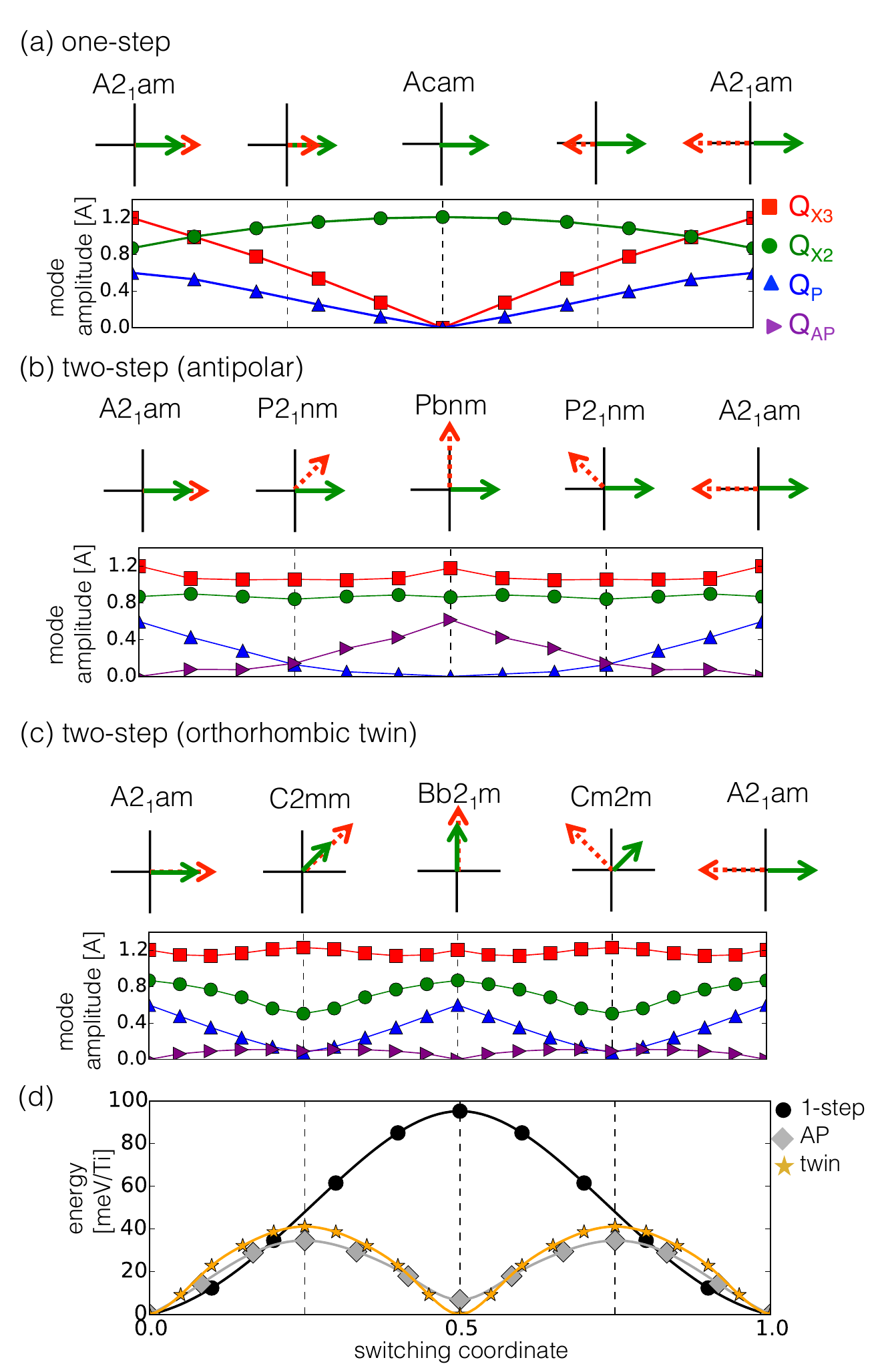}
 \caption{\label{x3-switching}  Ferroelectric switching pathways that reverse the $X_3^-$ octahedral rotation. (a) one-step switching, (b, c) two-step switching via antipolar ($Pbnm$) and orthorhombic twin ($Bb2_1m$) intermediate, respectively. The top of each panel shows the $\eta^{X_3^-}$ (red) and $\eta^{X_2^+}$ (green) OPs, while the bottom shows the amplitudes of all OPs. (d) The total energy as a function of switching coordinate for paths (a-c).}
 \end{figure}

Fig.~\ref{x3-switching} shows the paths that switch the polarization by reversing the $X_3^-$ octahedral rotation sense. These are analogous to the paths in Fig.~\ref{fig2} that reverse the $X_2^+$ rotation. In the one-step path, shown in Fig.~\ref{x3-switching}(a), $\eta^{X_3^-}$ turns off and then turns back on pointing in the opposite direction, the barrier has symmetry $Acam$.
In the antipolar path, shown in Fig.~\ref{x3-switching}(b), $\eta^{X_2^+}$ stays fixed, while  $\eta^{X_3^-}$ rotates, the intermediate phase is the antipolar orthorhombic twin $Pbnm$. Fig.~\ref{x3-switching}(c) shows switching via the $Bb2_1m$ orthorhombic twin intermediate, where both $\eta^{X_3^-}$ and $\eta^{X_2^+}$ rotate in the two dimensional OP space. 
 Fig.~\ref{x3-switching}(d) shows the total energy as a function of switching coordinate, we find that the two-step paths are both lower energy than the one-step path, with the antipolar path being the lowest. 

\section{\label{tri-app} Trilinear coupling}
Table~\ref{ops} presents the transformation properties of the OPs $\eta^{X_3^-}$, $\eta^{X_2^+}$, $\eta^P$, and $\eta^{AP}$ under the generators of $I4/mmm$. The generators of \ifour~ are: mirror plane $\sigma_z = (-x, -y, z)$, 90$^\circ$ rotation $C_{4z}^+ = (-y, x, z)$, mirror plane $\sigma_y = (-x,y,-z)$, inversion $I = (-x,-y,-z)$, and translations $\{E|\frac{1}{2}\frac{1}{2}\frac{1}{2}\} = (x+\frac{1}{2}, y+\frac{1}{2}, z+\frac{1}{2})$ and $\{E | 100\} = (x+1, y, z$). 

By inspecting Table~\ref{ops}, one can see that the bilinear $Z_1$ ($Z_2)$ has the same transformation properties as \npt~ (\npo) so $Z_1$\npt+$Z_2$\npo~ is invariant and thus is allowed in the free energy. Similarly, the bilinear $Z_3$ ($Z_4$) has the same transformation properties as \napt~ (\napo) so $Z_3$\napt+$Z_4$\napo~ is invariant  and allowed in the free energy. The sum of these are the four trilinear coupling terms presented in Eq.~\ref{eq:tri-2d}.

 \begin{table}
\begin{tabular}{c | c | c | c | c | c | c}
\hline
 & $\sigma_z$ & $C_{4z}^+$ & $\sigma_y$ & $I$ & $\{E|\frac{1}{2}\frac{1}{2}\frac{1}{2}\} $ & $\{E | 100\}$  \\
\hline
\ntho & -\ntho & \ntht & \ntht & -\ntho & \ntho & -\ntho \\
\ntht & -\ntht & -\ntho & \ntho & -\ntht & -\ntht & -\ntht \\
\hline
\ntwo & \ntwo & -\ntwt & \ntwt & \ntwo & \ntwo & -\ntwo \\
\ntwt & \ntwt & -\ntwo & \ntwo & \ntwt & -\ntwt & -\ntwt \\
\hline
\npo & -\npo & \npt & \npt & -\npo & \npo & \npo \\
\npt & -\npt & -\npo & \npo & -\npt & \npt & \npt \\
\hline
\napo & -\napo & \napt & \napt & -\napo & -\napo & \napo \\
\napt & -\napt & -\napo & \napo & -\napt & -\napt & \napt \\
\hline
\yo & -\yo & -\yt & \yt & -\yo & \yo & \yo \\
\yt & -\yt & \yo & \yo & -\yt & \yt & \yt \\
\zo & -\zo & -\zt & \zt & -\zo & -\zo & \zo \\
\zt & -\zt & \zo & \zo & -\zt & -\zt & \zt \\
\hline
\end{tabular}
\caption{\label{ops}Transformation properties of OPs $\eta^Y$, $Y=\{X_3^-, X_2^+, P, AP\}$ under generators of $I4/mmm$. Transformation properties of bilinears $Z_1=$\ntho\ntwo, $Z_2=$\ntht\ntwt, $Z_3=$\ntho\ntwt, and $Z_4=$\ntht\ntwo are also shown.}

\end{table}

Table~\ref{invariants} presents the transformation properties of the local OPs $T_\alpha$, $\Phi_\alpha$, and $P_\alpha$ ($\alpha=1,2$) under the generators of $I4/mmm$, as well as the transformation properties of the bilinears $Y_\alpha = P_{\alpha x}T_{\alpha y}-P_{\alpha y}T_{\alpha x}$. From Table~\ref{invariants}, it is clear that $Y_\alpha$ and $\Phi_\alpha$ have the same transformation properties, so the allowed trilinear terms are $Y_1\Phi_1 + Y_2\Phi_2$, which is the expression for $\mathcal{F}_\mathrm{tri}$ given in Eq.~\ref{eq:local_tri_1}.

 \begin{table}
\begin{tabular}{c | c | c | c | c | c | c}
\hline
 & $\sigma_z$ & $C_{4z}^+$ & $\sigma_y$ & $I$ & $\{E|\frac{1}{2}\frac{1}{2}\frac{1}{2}\} $ & $\{E | 100\}$  \\
\hline
$T_{1x}$ & -$T_{1x}$ &-$T_{1y}$ & -$T_{1y}$& -$T_{1x}$ & $T_{2x}$ & -$T_{1x}$ \\
$T_{1y}$ & -$T_{1y}$ & $T_{1x}$ & -$T_{1x}$ & -$T_{1y}$ & $T_{2y}$ & -$T_{1y}$ \\
$T_{2x}$ & -$T_{2x}$ &$T_{2y}$ & $T_{2y}$& -$T_{2x}$ & $T_{1x}$ & -$T_{2x}$ \\
$T_{2y}$ & -$T_{2y}$ & -$T_{2x}$ & $T_{2x}$ & -$T_{2y}$ & $T_{1y}$ & -$T_{2y}$ \\
\hline
$\Phi_1$ & $\Phi_1$ & -$\Phi_1$ & $\Phi_1$ & $\Phi_1$ &$\Phi_2$ & -$\Phi_1$ \\
$\Phi_2$ &$\Phi_2$& $\Phi_2$ & -$\Phi_2$ & $\Phi_2$ & -$\Phi_1$& -$\Phi_2$ \\
\hline
$P_{1x}$ & -$P_{1x}$ &$P_{1y}$ & $P_{1y}$& -$P_{1x}$ & $P_{2x}$ & $P_{1x}$ \\
$P_{1y}$ & -$P_{1y}$ & -$P_{1x}$ & $P_{1x}$ & -$P_{1y}$ & $P_{2y}$ & $P_{1y}$ \\
$P_{2x}$ & -$P_{2x}$ &$P_{2y}$ & $P_{2y}$& -$P_{2x}$ & $P_{1x}$ & $P_{2x}$ \\
$P_{2y}$ & -$P_{2y}$ & -$P_{2x}$ & $P_{2x}$ & -$P_{2y}$ & $P_{1y}$ & $P_{2y}$ \\
\hline
$Y_1$ & $Y_1$ & -$Y_1$ & $Y_1$& $Y_1$& $Y_2$ & -$Y_1$ \\
$Y_2$ & $Y_2$ & $Y_2$ & -$Y_2$& $Y_2$& $Y_1$ & -$Y_2$ \\
\hline
\end{tabular}
\caption{\label{invariants}Transformation properties of local OPs and bilinears $Y_\alpha = P_{\alpha x}T_{\alpha y}-P_{\alpha y}T_{\alpha x}$ $(\alpha = 1,2$) under generators of $I4/mmm$.  }

\end{table}


\section{\label{local-app} Relationship between the two sets of order parameters}


\begin{figure}
 \includegraphics[width=0.45\textwidth]{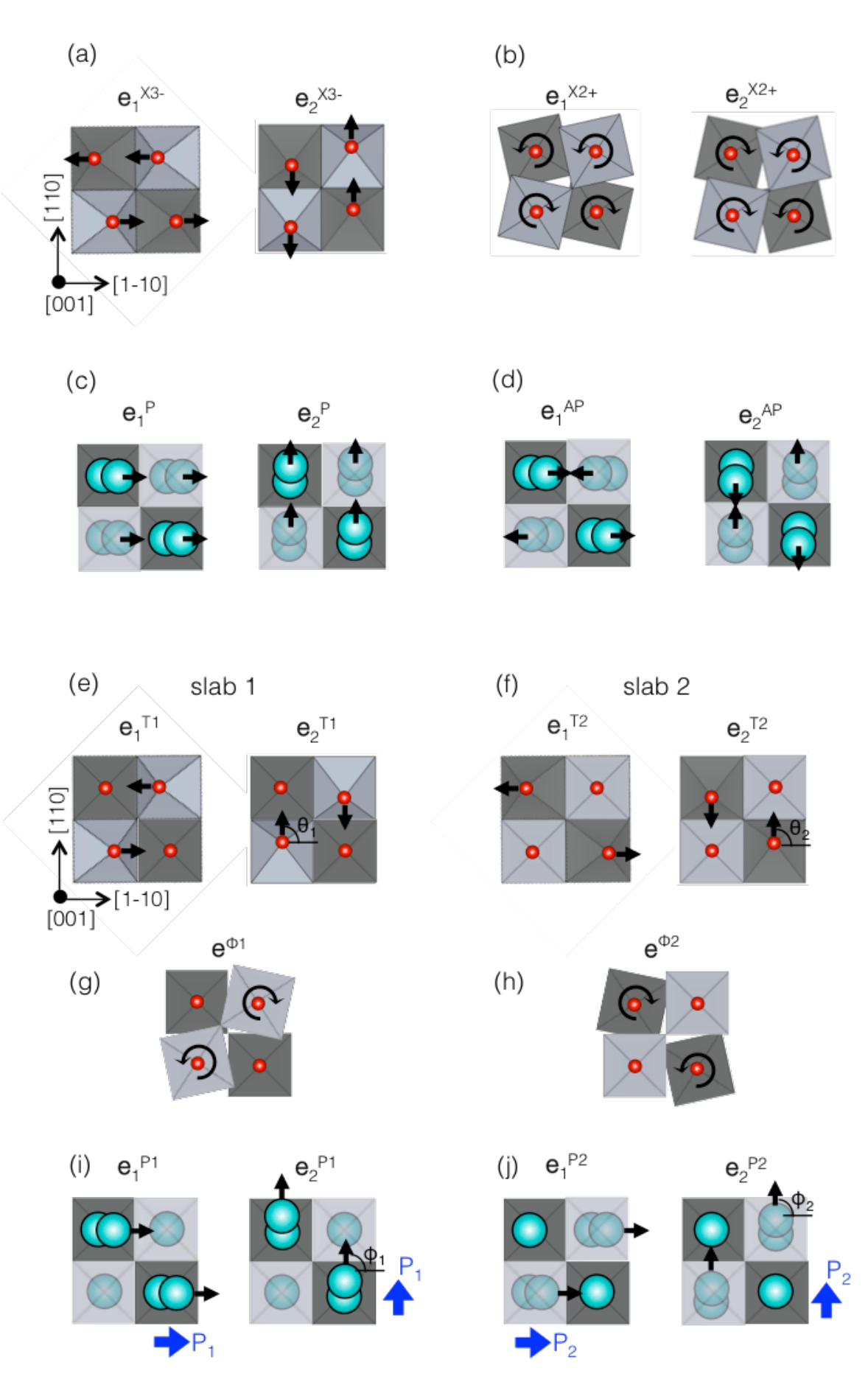}
 \caption{\label{i4mmm-basis} Basis modes for structural distortions that transform like irreps of $I4/mmm$ (a-d), and for local structural distortions described by local OPs (e-j). }
 \end{figure}

This appendix derives the relationship between the local OPs and those that transform like irreps of $I4/mmm$. This will allow us to show that the two expressions for $\mathcal{F}_\mathrm{tri}$ presented in Eqs.~\ref{eq:tri-2d} and ~\ref{eq:local_tri_1} are equivalent.
All structures considered in this work can be decomposed into distortions that transform like the irreps of $I4/mmm$:
\begin{equation}
{\bf u} = {\bf u}^{X_3^-} + {\bf u}^{X_2^+} + {\bf u}^{P} + {\bf u}^{AP}. \nonumber
\end{equation}
Here ${\bf u}$ is a vector that contains the atomic displacements of all atoms in the unit cell. A distortion that transforms like an arbitrary direction of irrep $Y$ described by OP ($\eta_1^Y$,$\eta_2^Y$) can be written as a linear combination of orthonormal basis distortion modes~\cite{perez-mato10} $\hat{\bf e}_{1,2}^Y$: ${\bf u}^Y = \eta_1^Y\hat{\bf e}_{1}^Y+\eta_2^Y\hat{\bf e}_{2}^Y$. These basis modes are shown in Fig.~\ref{i4mmm-basis}. Note that the basis modes for $X_3^-$ and $X_2^+$ in Fig.~\ref{i4mmm-basis}(a,b) are the $(a,0)$ and $(0,a)$ domains which were already shown in Fig.~\ref{struc}, but they are reproduced here for clarity of the present discussion. The basis modes for the polar and antipolar irreps, shown in Fig.~\ref{i4mmm-basis}(c,d) are polar (antipolar) displacements along [110] and [-110], respectively.

Equivalently, as discussed in the main text, each ${\bf u}^Y$ can be written as the sum of two distortions local to each perovskite slab $\alpha=1,2$: ${\bf u}^Y = \sum_{\alpha=1,2}{\bf u}^{(\alpha)}$. These local distortions ${\bf u}^{(\alpha)}$ are written in terms of their own basis modes ${\bf e}_{1,2}^{(\alpha)}$, shown in Fig.~\ref{i4mmm-basis}(e-j) ( already presented in Fig.~\ref{fig3}, but reproduced here). Here, for each irrep, we write down the relation between these two sets of basis modes, which will allow us to relate Eqs.~\ref{eq:tri-2d} and ~\ref{eq:local_tri_1}.

\subsection{$X_3^-$ octahedral tilting distortion}
Let ${\bf u}^{X_3^-}$ be a distortion described by OP $(\eta_1^{X_3^-},\eta_2^{X_3^-})$, it can be written as  
\begin{equation}
{\bf u}^{X_3^-} = \eta_1^{X_3^-}\hat{\bf e}_1^{X_3^-} + \eta_2^{X_3^-}\hat{\bf e}_2^{X_3^-} \nonumber
\end{equation}
where normalized basis modes $\hat{\bf e}_1^{X_3^-}$ and $\hat{\bf e}_2^{X_3^-}$ are shown in Fig.~\ref{i4mmm-basis}(a).

We can also write ${\bf u}^{X_3^-}={\bf u}^{(1)} + {\bf u}^{(2)}$, where ${\bf u}^{(\alpha)}$ is an out-of-phase tilting distortion in slab $\alpha$ described by  two dimensional OP $T_\alpha=(T_{\alpha x}, T_{\alpha y})$. It can be written as the sum of the two  basis modes $\hat{\bf e}_{1,2}^{(T\alpha)}$ shown in Fig.~\ref{i4mmm-basis}(e,f):
\begin{eqnarray}
{\bf u}^{(1)} &=& T_{1x} \hat{\bf e}_1^{(T1)} + T_{1y} \hat{\bf e}_2^{(T1)}  \nonumber \\
{\bf u}^{(2)} &=& T_{2x} \hat{\bf e}_1^{(T2)} + T_{2y} \hat{\bf e}_2^{(T2)}  \nonumber
\label{eq1}
\end{eqnarray}
These two sets of basis distortions are related as follows:
\begin{eqnarray}
\hat{\bf e}_1^{X_3^-} &=& (\hat{\bf e}_1^{(T1)} +  \hat{\bf e}_1^{(T2)})/\sqrt{2} \nonumber \\
\hat{\bf e}_2^{X_3^-} &=& (-\hat{\bf e}_2^{(T1)} +  \hat{\bf e}_2^{(T2)})/\sqrt{2}, \nonumber
\end{eqnarray}
so we can write 
\begin{equation}
{\bf u}^{X_3^-} = \frac{\eta_1^{X_3^-}}{\sqrt{2}} (\hat{\bf e}_1^{(T1)} +  \hat{\bf e}_1^{(T2)}) + \frac{\eta_2^{X_3^-} }{\sqrt{2}}(-\hat{\bf e}_2^{(T1)} +  \hat{\bf e}_2^{(T2)}). \nonumber
\end{equation}
Comparing  the expressions above, one can see that  $T_{1x} = T_{2x} = \eta_1^{X_3^-}/\sqrt{2}$, $-T_{1y} = T_{2y} = \eta_2^{X_3^-}/\sqrt{2}$, so
\begin{eqnarray}
\label{eq:t}
T_1 &=& \frac{1}{\sqrt{2}}(\eta_1^{X_3^-}, -\eta_2^{X_3^-})  \nonumber \\
T_2 &=& \frac{1}{\sqrt{2}}(\eta_1^{X_3^-},\eta_2^{X_3^-}).
\end{eqnarray}

\subsection{$X_2^+$ octahedral rotation distortion}
Here we consider distortion ${\bf u}^{X_2^+}$ described by OP $(\eta_1^{X_2^+},\eta_2^{X_2^+})$, 
\begin{equation}
{\bf u}^{X_2^+} = \eta_1^{X_2^+} \hat{\bf e}_1^{X_2^+} + \eta_2^{X_2^+}\hat{\bf e}_2^{X_2^+} \nonumber
\end{equation}
where the basis modes $\hat{\bf e}_{1,2}^{X_2^+}$ are shown in Fig.~\ref{i4mmm-basis}(b).

 Also, we can write ${\bf u}^{X_2^+}={\bf u}^{(1)} + {\bf u}^{(2)}$ as the sum of $a^0a^0b^+$ in-phase rotation distortions ${\bf u}^{(\alpha)}$ local to each slab $\alpha$ described by local OP $\Phi_\alpha$.  The local distortions can be written as ${\bf u}^{(\alpha)} = \Phi_\alpha \hat{\bf e}^{(\Phi\alpha)}$, where the basis modes $\hat{\bf e}^{(\Phi\alpha)}$ are shown in Fig.~\ref{i4mmm-basis}(g,h). The relationship between the two sets of basis modes is:
\begin{eqnarray}
\hat{\bf e}_1^{X_2^+}& =&  (-\hat{\bf e}^{(\Phi 1)} -  \hat{\bf e}^{(\Phi 2)})/\sqrt{2}  \nonumber \\
\hat{\bf e}_2^{X_2^+} &=& (-\hat{\bf e}^{(\Phi 1)} +  \hat{\bf e}^{(\Phi 2)})/\sqrt{2}, \nonumber
\end{eqnarray}
so
\begin{equation}
{\bf u}^{X_2^+} = \frac{\eta_1^{X_2^+}}{\sqrt{2}}( -\hat{\bf e}^{(\Phi 1)} -  \hat{\bf e}^{(\Phi 2)}) + \frac{\eta_2^{X_2^+}}{\sqrt{2}}(-\hat{\bf e}^{(\Phi 1)} +  \hat{\bf e}^{(\Phi 2)}). \nonumber
\end{equation}
Therefore, the relationship between the two sets of OPs is:
\begin{eqnarray}
\Phi_1 &=& -\frac{1}{\sqrt{2}}(\eta_1^{X_2^+} + \eta_2^{X_2^+}) \nonumber \\
\Phi_2 &=& -\frac{1}{\sqrt{2}}( \eta_1^{X_2^+} - \eta_2^{X_2^+}). 
\label{eq:r}
\end{eqnarray}

\subsection{Polar (P) and antipolar (AP) distortions ($\Gamma_5^-$ and $M_5^-$)}
Finally, we consider distortions with polar and antipolar OPs $(\eta_1^P,\eta_2^P)$ and $(\eta_1^{AP},\eta_2^{AP})$:
\begin{eqnarray}
{\bf u}^P &=& \eta_1^P\hat{\bf e}_1^P + \eta_2^P\hat{\bf e}_2^P  \nonumber \\
{\bf u}^{AP} &=& \eta_1^{AP}\hat{\bf e}_1^{AP} + \eta_2^{AP}\hat{\bf e}_2^{AP} \nonumber
\end{eqnarray}
where the basis modes are shown in Fig.~\ref{i4mmm-basis}(c,d). 
As discussed in the main text, polar distortions local to each slab ${\bf u}^{(\alpha)}$ are described by the two-dimensional local OP $P_\alpha=(P_{\alpha x}, P_{\alpha y})$, and can be written as the sum of the basis modes shown in Fig.~\ref{i4mmm-basis}(i,j): ${\bf u}^{(\alpha)} = P_{\alpha x} \hat{\bf e}_1^{(P\alpha)} + P_{\alpha y} \hat{\bf e}_2^{(P\alpha)}$. 
The two sets of basis distortions are related as follows:
\begin{eqnarray}
\hat{\bf e}_1^{P} &=& (\hat{\bf e}_1^{(P1)} +  \hat{\bf e}_1^{(P2)})/\sqrt{2}  \nonumber \\
\hat{\bf e}_2^{P} &=& (\hat{\bf e}_2^{(P1)} +  \hat{\bf e}_2^{(P2)})/\sqrt{2} \nonumber
\end{eqnarray}
and
\begin{eqnarray}
\hat{\bf e}_1^{AP} &=& (\hat{\bf e}_1^{(P1)} -  \hat{\bf e}_1^{(P2)})/\sqrt{2} \nonumber \\
\hat{\bf e}_2^{AP} &=& (\hat{\bf e}_2^{(P1)} -  \hat{\bf e}_2^{(P2)})/\sqrt{2}. \nonumber
\end{eqnarray}

Therefore,
\begin{eqnarray}
{\bf u}^{P} &=& \frac{\eta_1^{P}}{\sqrt{2}}( \hat{\bf e}_1^{(P1)} +  \hat{\bf e}_1^{(P2)}) + \frac{\eta_2^{P}}{\sqrt{2}}(\hat{\bf e}_2^{(P1)} +  \hat{\bf e}_2^{(P2)}) \nonumber \\
{\bf u}^{AP} &=& \frac{\eta_1^{AP}}{\sqrt{2}}( \hat{\bf e}_1^{(P1)} -  \hat{\bf e}_1^{(P2)}) + \frac{\eta_2^{AP}}{\sqrt{2}}(\hat{\bf e}_2^{(P1)} -  \hat{\bf e}_2^{(P2)}).  \nonumber 
\end{eqnarray}

We can write the total distortion from both modes as ${\bf u} = {\bf u}^P + {\bf u}^{AP} = {\bf u}^{(1)} + {\bf u}^{(2)}$, so 
\begin{eqnarray}
{\bf u}^{(1)} &=&  \frac{1}{\sqrt{2}}(\eta_1^P + \eta_1^{AP})\hat{\bf e}_1^{(P1)} +  \frac{1}{\sqrt{2}}(\eta_2^P + \eta_2^{AP})\hat{\bf e}_2^{(P1)} \nonumber \\ \nonumber
{\bf u}^{(2)} &=&  \frac{1}{\sqrt{2}}(\eta_1^P - \eta_1^{AP})\hat{\bf e}_1^{(P2)} +  \frac{1}{\sqrt{2}}(\eta_2^P - \eta_2^{AP})\hat{\bf e}_2^{(P2)}.
\end{eqnarray}

From which we can obtain the relation:
\begin{eqnarray}
P_1 &=& \frac{1}{\sqrt{2}}(\eta_1^P + \eta_1^{AP}, \eta_2^P + \eta_2^{AP})  \nonumber \\
P_2 &=&  \frac{1}{\sqrt{2}}(\eta_1^P - \eta_1^{AP}, \eta_2^P - \eta_2^{AP}).
\label{eq:p}
\end{eqnarray}

Using the relations between the two sets of OPs given in Eq.~\ref{eq:t}-\ref{eq:p}, one can show that the two expressions for $\mathcal{F}_\mathrm{tri}$ given in Eqs.~\ref{eq:tri-2d} and ~\ref{eq:local_tri_1} are equivalent. 

\bibliographystyle{apsrev}
\bibliography{GiantBibliography}

\begin{thebibliography}{45}
\bibitem[{foo({\natexlab{a}})}]{foot3}
\bibinfo{note}{For the $X_3^-$ and $X_2^+$ irreps, the OPs are two dimensional
  because there are two wavevectors in the star of $X$.}
\bibitem[{foo({\natexlab{b}})}]{foot1}
\bibinfo{note}{There are additional trilinear couplings to other modes when
  $\eta^{X_3^-}$ and/or $\eta^{X_2^+}$ are along a general direction in OP
  space, but we neglect them because the polar and antipolar modes are the ones
  that are physically relevant to this problem.}
\bibitem[{foo({\natexlab{c}})}]{foot7}
\bibinfo{note}{Note that the structure has symmetry $P2_1am$ at all points
  along the switching path except when the switching coordinate is 0, 0.5, and
  1, not just at the barrier.}
\bibitem[{foo({\natexlab{d}})}]{foot8}
\bibinfo{note}{When both OPs are along a general direction (all points on
  switching path except when switching coordinate = 0, 0.25, 0.5, 0.75, 1), the
  symmetry is $Pm$. The reason that the `barrier' structure is a different
  symmetry for the orthorhombic twin path and it is not for the antipolar path
  is because the orthorhombic path is connecting two domains of the same space
  group ($A2_1am$, $Bb2_1m$), while the antipolar path is connecting domains of
  two different space groups ($A2_1am$, $Pnam$).}
\end{thebibliography}

\end{document}